# Chemical Fractionation in the Silicate Vapor Atmosphere of the Earth

Kaveh Pahlevan*, David J. Stevenson and John M. Eiler
Division of Geological and Planetary Sciences
California Institute of Technology
MC 150-21, Pasadena, CA 91125
United States of America

## Abstract

Despite its importance to questions of lunar origin, the chemical composition of the Moon is not precisely known. In recent years, however, the isotopic composition of lunar samples has been determined to high precision, and found to be indistinguishable from the terrestrial mantle, despite widespread isotopic heterogeneity in the Solar System. In the context of the giant impact hypothesis, a high level of isotopic homogeneity can be generated if the proto-lunar disk and post-impact Earth undergo turbulent mixing into a single uniform reservoir while the system is extensively molten and partially vaporized. In the absence of liquid-vapor separation, such a model leads, in addition, to the lunar inheritance of the Earth mantle chemical composition. Hence, the turbulent mixing model raises the question of how chemical differences arose between the silicate Earth and Moon. Here we explore chemical and isotopic consequences of liquid-vapor separation in one of the settings relevant to the lunar composition: the silicate vapor atmosphere of the post giant-impact Earth. We use a model atmosphere to quantify the extent to which rainout in convective clouds can generate chemical differences by enriching the upper atmosphere in the vapor, and show that plausible parameters can generate the postulated ~2x enhancement in the FeO/MgO ratio of the silicate Moon relative to the Earth's mantle. Moreover, we show that liquid-vapor separation also generates measurable mass-dependent isotopic offsets between the silicate Earth and Moon even at the high temperatures encountered after the giant impact. Finally, we use the concomitant nature of chemical and isotopic fractionation and the absence of a silicate Earth-Moon difference in the isotopic composition of silicon to conclude that either the major-element silicate Moon is nearly isochemical with the terrestrial mantle or that the major-element composition of the Moon was influenced by liquid-vapor separation in a setting other than the silicate vapor atmosphere of Earth. An approach of this kind has the potential to resolve long-standing questions on the lunar chemical composition.

*Corresponding Author: kaveh@gps.caltech.edu

## 1. Introduction

The Moon is generally thought to have formed from the debris of a massive, off-center collision of the proto-Earth with a smaller planet towards the end of accretion (Cameron and Ward 1976, Hartmann and Davis 1975). Such a scenario can explain the angular momentum of the Earth-Moon system (Cameron and Ward 1976), the lunar metal depletion (Cameron 2000, Canup 2004b, Canup 2008, Canup and Asphaug 2001), the small, but non-zero inclination of the lunar orbit (Ward and Canup 2000) and the terrestrial isotopic character of the lunar material (Pahlevan and Stevenson 2007, Wiechert et al. 2001). However, no quantitative model exists that can explain the chemical composition of the silicate Moon and its relationship to high-temperature processes that must have followed a giant impact.

A weakness of the giant impact hypothesis has been that its predictions regarding lunar chemistry have never been clearly articulated. The reasons are two-fold. First, smooth-particle hydrodynamic (SPH) impact simulations (Cameron 2000, Canup 2004b, Canup 2008, Canup and Asphaug 2001) consistently derive the initial disk of material orbiting the Earth primarily from the mantle of the impactor whose composition cannot be independently known. Second, we do not yet have a complete scenario of lunar formation starting with the impact event and ending with a fully formed Moon (Canup 2004a, Stevenson 1987). In particular, the SPH simulations generate a hot, melt-vapor circumterrestrial disk on a timescale of hours while it may take ~$10^3$ years of cooling before lunar accretion sets in, a time during which the lunar-forming material may have been processed and altered. The compositional signatures of the giant impact may have been imprinted while the lunar material existed in the form of such a melt-vapor disk (Machida and Abe 2004, Thompson and Stevenson 1988) or even during lunar accretion (Abe et al. 1998), important stages of the evolution – after the impact event but prior to the formation and closure of the Moon – that have received limited attention. The giant impact hypothesis cannot be tested via comparison with cosmochemical observations without an articulation of the consequences of these stages for the lunar composition.

The passage of the lunar-forming material through such a high-temperature, partially-vaporized disk makes possible isotopic homogenization of the terrestrial magma ocean with the proto-lunar magma disk through turbulent mixing while the system exists in the form of a continuous fluid (Pahlevan and Stevenson 2007). A growing body of evidence is reconciled through such an episode of thorough, system-wide turbulent mixing in the $\sim 10^3$ years following the giant impact. First, the oxygen isotopic composition of the Moon is identical with the terrestrial mantle to within $< 5$ ppm (Wiechert et al. 2001) while the composition of the Moon-forming impactor was very likely distinct from the terrestrial-lunar value at this level of precision (Pahlevan and Stevenson 2007). The Earth-Mars difference, for comparison, is ~320 ppm (Franchi et al. 1999). Second, the silicon isotope composition of the silicate Earth has been measured to high precision, and found to be distinct from the chondritic meteorites (Fitoussi et al. 2009, Georg et al. 2007), an offset that is thought to reflect isotopic fractionation accompanying silicon segregation into the core. Because silicon dissolution into core-bound metal requires high-temperatures and pressures (Gessmann et al. 2001), this process is thought possible for the Earth, but not the Moon-forming impactor. In this context, the silicon isotope composition of the Moon, distinct from primitive meteorites but indistinguishable from the silicate Earth (Armytage et al. 2010, Fitoussi et al. 2010, Georg et al. 2007) is unexplained without an episode of mixing to homogenize the isotopes of silicon in the fluid aftermath of the giant impact. Third, new measurements of tungsten isotopes in lunar metals – largely impervious to cosmic ray signal degradation – have revealed the W isotope composition of the silicate Moon to be indistinguishable from the Earth's mantle (Touboul et al. 2007, Touboul et al. 2009). Because tungsten isotopes measure the rate of accretion and core formation, and because the accretion of smaller terrestrial planets is generally faster than the protracted accretion of larger terrestrail planets (Kleine et al. 2002), and because of the asymmetric nature of the giant impact, with a mass ratio of ~1:9 (Canup and Asphaug 2001) or ~2:8 (Canup 2008), tunsten isotopic uniformity between the silicate Earth and the lunar disk is unexpected without appealing to an episode of large-scale mixing between these reservoirs (Nimmo et al. 2010). Finally, giant impact simulations that include pre-impact rotation have, for the first time, been carried out (Canup 2008). Despite such an advance, the conclusion from previous studies that the material initially injected into proto-lunar orbit is predominantly sourced from the mantle of the impactor stands. The turbulent mixing hypothesis is unique among giant impact scenarios in that it derives the lunar material from the Earth's mantle. Moreover, this hypothesis is attractive because it has testable consequences. The goal of this paper is to develop some of these consequences.

One consequence of the turbulent mixing hypothesis is that liquid-vapor equilibrium and partial separation may have played a central role in lunar formation. At sufficiently high temperature, the isotopic fractionation between co-existing phases approaches zero (Urey 1947). Here, we show that – even at the temperatures prevailing after the giant impact – measurable isotopic fractionation between the liquid and vapor may occur. Chemically, because silicate mantles are multi-component systems, the composition of a silicate liquid and its co-existing vapor will, in general, be distinctly different. This behavior of high temperature equilibrium – of similar isotopic but different chemical composition of co-existing phases – makes it a prime candidate for explaining isotopic similarities and chemical differences in the silicate Earth-Moon system. Below, we explore a scenario in which liquid-vapor fractionation in the post giant-impact environment can evolve chemical (and small, mass-dependent isotopic) differences even as it eliminates mass-independent isotopic heterogeneity via large-scale mixing.

## 2. The Role of Liquid-Vapor Fractionation

There are multiple settings in which liquid-vapor fractionation may have taken place, each with different consequences for the composition of the Moon. First, there is phase separation (rainout) in the Earth's post-impact atmosphere. This is relevant to the lunar composition if the proto-lunar disk exchanges material with the Earth's silicate vapor atmosphere and ultimately inherits its composition, as has been proposed to account for the terrestrial isotopic character of lunar matter (Pahlevan and Stevenson 2007). In such a setting, any fractionation leads to vapor enrichment in the lunar composition, as rainout removes droplets from ascent to the upper atmosphere that Earth exchanges with the proto-lunar disk. Second, fractionation is possible in the proto-lunar disk itself, the precipitation of droplets towards the mid-plane separating the liquid from the vapor.

However, due to our poor understanding of the disk evolution, we do not yet know if phase-separation in the disk ultimately enhances the lunar material in the liquid or vapor. Finally, fractionation is possible via outgassing from a lunar magma ocean and escape of the resulting transient atmosphere from the weak lunar gravity field, or from precursor liquid moonlets. Such fractionation would enhance the lunar material in liquid through the elimination of vapor.

The lunar bulk composition is not well known. There are mutually exclusive viewpoints on the bulk lunar FeO/MgO ratio and refractory element abundances, for example. One of the goals of this work is to calculate the range of lunar compositions that can be derived from the Earth mantle via liquid-vapor fractionation. As we show below, the behavior of the most abundant components (FeO-MgO-$SiO_2$) is tied to the fate of more refractory (i.e. CaO, $Al_2O_3$) as well as more volatile (i.e. $Na_2O$, $K_2O$) elements. Partitioning Earth mantle material into a liquid and complementary vapor, a single stage enhancement of vapor via partial removal of liquid leads to a composition with a higher FeO/MgO ratio, a depletion of refractory elements, and an enhancement in volatile elements. Hence, a one-stage model producing a higher lunar FeO/MgO ratio (discussed below) is incapable of generating an enrichment of refractory elements or the volatile element depletion of the lunar material (Ringwood and Kesson 1977) starting from Earth mantle composition.

Large depletions of volatile elements may, nevertheless, occur via partial vaporization or incomplete condensation and removal of the most volatile fraction of the material, for example, during the lunar magma ocean and molten disk stage, respectively. Since extreme depletions of volatile elements are observed (Ringwood and Kesson 1977), such an episode of fractionation must be taken seriously. However, the process responsible for the volatile-element depletion may have been de-coupled from the major-element fractionation, leaving the lunar FeO/MgO ratio essentially unchanged. Support for this proposition comes from the fact that no researcher has claimed that the silicate Moon is *depleted* in its FeO content relative to Earth mantle. Hence, if the lunar material was derived from the silicate Earth via liquid-vapor fractionation, then the process responsible for the volatile depletion was evidently not effective at depleting the most volatile major-element component, FeO. The volatile depletion may have been restricted to elements less abundant than the major elements for several reasons, for example, if the process responsible for the loss was energy-limited.

Here, we restrict ourselves to fractionation that occurs via rainout in the post giant-impact silicate atmosphere of Earth. We focus on a major-element compositional feature: the FeO/MgO ratio of the silicate Moon, a ratio sometimes inferred to be significantly higher than (Khan et al. 2007, Ringwood 1977) and sometimes inferred to be nearly identical with (Warren 1986, Warren 2005) the terrestrial mantle. We choose this chemical tracer for several reasons. First, both iron and magnesium are major elements in silicate mantles and their lunar abundances are therefore constrained, even in unsampled regions, by geophysical measurements. Second, these oxides – and their corresponding silicates – have significantly different behavior during vaporization, FeO being much more volatile than MgO (Nagahara et al. 1988), making the FeO/MgO ratio a sensitive tracer of liquid-vapor separation. Third, thermodynamic data for major elements exist at the temperatures of relevance (T = 2,500 – 3,500 K). The question we ask is whether an elevated lunar FeO/MgO ratio can be evolved from the partially vaporized magma ocean of Earth via rainout in the surrounding vapor atmosphere and what the compositional consequences of this process would be. Since the isotopic composition of the liquid and vapor will, in general, be different, rainout is also expected to imprint mass-dependent isotopic signatures. The principle conclusion of this paper is that a range of compositions can evolve from the terrestrial magma ocean, and that chemical and isotopic fractionation are concomitant. We quantify the co-evolution of the FeO/MgO and $^{30}Si/^{28}Si$ ratio from the silicate Earth value due to varying degrees of liquid-vapor fractionation and use existing isotopic measurements on lunar and terrestrial samples to determine the range of lunar bulk compositions consistent with the turbulent mixing scenario here explored (and further discussed in section 5).

**3. Mantle Model**

The composition of the Earth's mantle is constrained by direct samples from the mantle, the composition of basalts generated by partial melting of the mantle interpreted with the insights of experimental petrology, and seismology. Through these methods, it has been determined that the Earth's upper mantle has an FeO content of ~8 weight percent and a molar Fe/Fe+Mg ratio of ~0.1 (O'Neill and Palme 1998), as well as an Mg/Si ratio that places its composition (in the MgO-$SiO_2$-FeO system) between olivine $(Fe,Mg)_2SiO_4$ and pyroxene $(Fe,Mg)SiO_3$. A compositional difference between the upper mantle and the deep mantle cannot be excluded but is not compelled by any geochemical or geophysical argument. Because we are interested in investigating the FeO/MgO ratio, we approximate the terrestrial mantle as a solution of ferromagnesian olivine with an Fe/Fe+Mg ratio of ~0.1.

Determining the lunar bulk composition is more difficult. The Moon is a highly differentiated object and unambiguously determining its composition from samples is not possible. There are several reasons for this. First, the crust is a major reservoir for some elements (i.e. aluminum) such that uncertainties in its thickness and compositional stratification result in major uncertainty in determining the lunar bulk composition (Taylor et al. 2006). Second, post-formation mixing of the lunar interior via mantle convection has been weak or absent, as evinced by the radiogenic isotopic heterogeneity (i.e. Sr isotopes) of samples derived from distinct regions of the lunar mantle by partial melting (Turcotte and Kellogg 1986). Such isotopic heterogeneity implies that the chemical heterogeneity generated during the primordial lunar differentiation has also survived. Finally, it is not yet clear how deep the reservoir is from which surface samples are derived and the deep lunar mantle, in particular, may be unsampled.

Nevertheless, petrologic studies have concluded that the source region for mare basalts is significantly enriched in FeO relative to the terrestrial mantle (Ringwood 1977). Geophysical measurements have also been brought to bear on the question. In particular, interior models must simultaneously satisfy the density, moment of inertia, and seismic velocity constraints. Geophysical constraints have tended to corroborate the suggestion from lunar petrology that the lunar FeO content is somewhat elevated relative to Earth mantle (Hood and Jones 1987, Khan et al. 2007, Mueller et al. 1988). However, the uncertainties are large, and estimates for the bulk lunar FeO content and FeO/MgO ratio range from values close to the Earth mantle (Warren 1986, Warren 2005) to significant enhancements (~13% FeO by weight, Fe/Fe+Mg ~ 0.2) relative to the silicate Earth (Taylor et al. 2006). Both compositions can be reproduced in the context of the present model, albeit with different consequences for the isotopic composition of lunar matter, a circumstance that enables the chemical composition of the Moon to be constrained by measurement its isotopic composition (section 7). We model the silicate Moon as an olivine solution with an Fe/Fe+Mg ratio of 0.1-0.2. The question we ask is under what conditions can such a composition evolve via fractionation from the post giant-impact silicate Earth, and what the consequences for the lunar isotopic composition would be.

**4. Thermodynamic Model**

The low-pressure form of the most abundant silicate mineral in the terrestrial planets is olivine $(Mg,Fe)_2SiO_4$. For the purposes of this study, we have chosen to take this mineral composition to represent the silicates, the only compositional parameter being the Fe/Mg ratio. Experiments have shown that olivine, as liquid or crystal, evaporates congruently (Mysen and Kushiro 1988, Nagahara et al. 1988), at least at the temperatures relevant to the solar nebula ($T < 2{,}000$ K). The congruence refers to the fact that two di-valent cations (iron or magnesium) enter the vapor phase for each silicon atom, but the vapor, despite having olivine composition, has an Fe/Mg ratio different than the coexisting condensate. Here, we assume that this vaporization behavior of olivine extends to the higher temperatures encountered on a post giant-impact Earth, and that the only components are the olivine end-members. This assumption can be tested by the extension of thermodynamic models of multi-component liquids (Ghiorso et al. 2002, Ghiorso and Sack 1995) to the temperatures of relevance (T = 2,500-3,500 K), or by performing an experiment in this temperature range.

We consider only two phases: liquid and vapor. The temperatures here encountered are above the liquidus temperature of olivine (2160 K). Solids are therefore assumed to be absent. For the liquid, we assume an ideal solution between olivine end-members. We take the vapor to be a mixture of ideal gases, and include atomic, monoxide, and dioxide metal-bearing species, as well as atomic and molecular oxygen (see Table 2). The thermodynamic data for phase equilibrium calculations come from standard compilations (Chase et al. 1985, Robie et al. 1978). Following (Fegley and Cameron 1987), we fit the standard state Gibbs free energy of reaction with a constant entropy and enthalpy. In some cases, this procedure represents a significant approximation. The thermodynamic data for liquid fayalite, in particular, only exists up to 1,800 K and its application here therefore requires a considerable extrapolation. A summary of the data used is given in Table 1. For evaporation-condensation reactions, for example, of forsterite, we have:

$$2MgSi_{1/2}O_2(l) = 2MgO(v) + SiO_2(v) \tag{1}$$

where the left hand side is equivalent to $Mg_2SiO_4(l)$, but is written such that the exchange of a single atom describes the difference between the liquid components. This choice of component makes it possible to describe the liquid as an ideal solution ($a_i = x_i$) where $a_i$ and $x_i$ are the activity and mole fraction of the ith component in the liquid (Navrotsky 1987). The statement of chemical equilibrium for a two-phase system is the equality of chemical potentials for each component in both phases. For the reaction above, the equilibrium constant can be written:

$$K_1 = \frac{P(MgO)^2 \times P(SiO_2)}{x_{Fo}^2} = \exp(-\Delta G_1^o / RT) \quad (2)$$

where we have additionally assumed that the vapor species behave as ideal gases ($f_i = P_i$) where $f_i$ and $P_i$ are the fugacity and partial pressure of the ith gaseous species. An equation of this kind is written for each reaction in Table 1, relating the partial pressures and liquid mole fractions to the standard-state Gibbs free energy of reactions. Knowledge of the congruent vaporization behavior of olivine is incorporated by requiring that the vapor have olivine composition, i.e. (Fe+Mg)/Si = 2, O/Si = 4.

$$P(FeO) + P(Fe) + P(MgO) + P(Mg) = 2 \times \sum_i P_i^{Si} \quad (3)$$

$$2P(O_2) + P(O) = 4 \times \sum_i P_i^{Si} \quad (4)$$

where $\sum P_i^{Si}$ is the sum over Si-bearing vapor species, $P(SiO_2) + P(SiO) + P(Si)$. The partial pressure of molecular oxygen is counted twice because we are counting atoms. By the Gibbs phase rule ($f = 2 + c - \phi$ where f is the number of independent variables, c is the number of components, and $\phi$ is the number of phases), a two-phase, two-component system is di-variant (f=2). The olivine end-members are the only components because we have incorporated the knowledge that olivine vaporizes congruently. However, specifying two parameters in a di-variant system allows one to solve for all the intensive properties (i.e. specific entropy, composition) of the phases present, not their relative amounts. Because we are interested in studying parcels with varying degrees of vapor, finding solutions requires that we specify an extra parameter (such as the specific entropy of the parcel) that constrains the degree of vaporization. This calculation is completely analogous to calculations of isentropic melting in the modern mantle (Asimow et al. 1997).

Parcel calculations require expressions for the specific entropy of the phases present, which – for an ideal solution and a mixture of ideal gases – can be related in a straightforward manner to the entropy of the constituents as pure substances. For an ideal liquid solution, we have:

$$\bar{S}_L(T,P) = \sum_i x_i \bar{S}_i(T,1) - 2R\sum_i x_i \ln x_i \tag{5}$$

where $S_L$ is the entropy of the liquid per mole of silicon atoms, $x_i$ is the mole fraction of the ith component in the melt, $S_i$ the entropy of the ith component as a pure liquid, and R the universal gas constant. The first term is a sum over the components as pure liquids, and the second term the entropy of mixing term, the factor of two resulting from the substitution of two atoms between olivine end-members. We assume that the pressure is sufficiently low as to not affect the liquid structure. Such an approximation should be fairly accurate since we are only concerned with two-phase regions of the Earth, and the calculations here indicate that the pressure below which vapor begins to co-exist with the liquid (<0.1 GPa) is much less than the bulk modulus of liquid silicates (~18 GPa). For a mixture of ideal gases, we have:

$$\bar{S}_v(T,P) = \left( \sum_i x_i \bar{S}_i(T,1) - R\sum_i x_i \ln x_i - R\ln P \right) \times \frac{1}{f_{Si}} \qquad (6)$$

with the first term in parenthesis the sum over species of the entropy of the gas molecules as pure gases at standard pressure (taking into account the entropy from rotational and vibrational degrees of freedom), the second term an entropy of mixing term arising from the fact that the vapor is a mixture of randomly distributed gas molecules, and the third term a pressure correction for ideal gases, which relates to the volume available to each molecule. The expression in parenthesis is the entropy per mole of gas molecule. Therefore, for consistency with the expression for the entropy of the liquid, we divide by $f_{Si}$, the fraction of gas molecules that are silicon-bearing ($f_{Si} = (P_{SiO2}+P_{SiO}+P_{Si})/P$) which results in an expression for the entropy of the vapor per mole of silicon atoms. The data used for the pure liquids and gases are listed in Table 2.

Finally, we assume that we are dealing with a pure system, i.e. no other species are present besides those accounted for explicitly. We therefore have expressions for the purity of the liquid and vapor solutions:

$$\sum_i x_{i,L} = 1 \qquad (7)$$

$$\sum_i P_i = P \quad (8)$$

For the liquid, the greatest correction to this model approximation comes from the presence of pyroxene (i.e. an (Fe+Mg)/Si ratio < 2) as well as the oxide components CaO and $Al_2O_3$ each of which have abundances of several percent. We later consider the fate of these refractory oxides under the assumption that they partition quantitatively into the liquid as passive tracers (section 6). For the vapor, the greatest impurity may be metallic iron – droplets from the impactor's core that remain suspended in the Earth's magma ocean after the giant impact, vaporizing along with the silicate liquid upon ascent. A potentially important correction for the vapor comes from the fact that we are using the ideal gas law, and near the critical point, the density of the vapor approaches the density of the liquid and intermolecular forces cannot be neglected. Such refinements will be the subject of future studies. Finally, closed-system behavior of parcels requires statements of mass balance and entropy additivity:

$$X = f_V X_V + (1 - f_V) X_L \quad (9)$$

$$\bar{S} = f_V \bar{S}_V + (1 - f_V) \bar{S}_L \quad (10)$$

where $f_v$ is the mole fraction of silicon atoms (or olivine units) that are vaporized, X and S are the composition (=Fe/Fe+Mg) and specific entropy of the atmospheric parcel, and $X_L$ and $X_V$ are the composition of the liquid and vapor phases, respectively. Equation (2) and six analogous equations (one for each reaction in Table 1), and equations (3)-(10) constitute the constraints in this problem. Once the thermodynamic data are specified (Tables 1 and 2), there are eighteen variables and fifteen constraints. We must therefore specify three variables to fully determine the system of equations: the parcel composition (X = 0.1), the parcel specific entropy (S = 10 $k_B$/atom) and the total pressure (P = 0.01-1 bar). By taking the composition of the well-mixed magma ocean to be equal to that of the modern mantle, we have implicitly assumed that the composition of the mantle has been unaltered since the Moon-forming giant impact. This assumption is discussed later (section 8.3). The chosen values for the specific entropy and total pressure are discussed in sections 5 and 6, respectively. We solve for the temperature (T), the vapor fraction ($f_V$), the specific entropy ($S_L$, $S_V$) and composition ($X_L$, $X_V$) of the liquid and vapor phases, and the partial vapor pressure of each of nine vapor species ($P_i$) by simultaneously solving the described system of coupled, non-linear equations using the MAPLE software package.

## 5. Thermal State

With a thermodynamic model of liquid-vapor silicates, we can construct one-dimensional models of the convective magma ocean of Earth and the transition to the overlying silicate vapor atmosphere under certain assumptions about the relevant dynamics. In the absence of phase changes, the energy released due to the impact of a ~0.1 Earth mass impactor onto the proto-Earth generates a system-wide mean temperature increase:

$$\Delta T = \frac{0.1GM_E}{R_E C_P} = 4{,}000\,K \quad (11)$$

with $C_p$ = 1,400 J/kg.K appropriate to the high-temperature, Dulong-Petit limit. For a ~0.2 Earth mass giant impactor (Canup 2008), the mean temperature rise is correspondingly twice as great. In all cases, however, vaporization contributes negligibly to the energy budget of Earth because – as discussed in section 4 – only the outer, low-pressure regions can undergo vaporization. For this reason, a mean temperature increase for the Earth of several thousand degrees is inevitable and it is nearly guaranteed that the post giant-impact silicate Earth is a magma ocean enveloped by a vapor atmosphere. Heat loss via radiation from a ~2,500 K photosphere (Thompson and Stevenson 1988) drives vigorous convection throughout the system.

We cannot calculate the chemical consequences of large-scale turbulent mixing without an understanding of the relevant dynamics in such a strongly driven system. One possibility is that of a convective magma ocean, a liquid-vapor interface, and a separately convective vapor atmosphere for both the Earth and the proto-lunar disk, with material exchange between the terrestrial and lunar reservoirs restricted to elements that partition into the vapor atmosphere. This was the picture originally discussed in connection with the turbulent mixing hypothesis (Pahlevan and Stevenson 2007). However, motivated, in part, by observations of refractory element isotopic homogeneity in the silicate Earth-Moon system (Touboul et al. 2007) and, in part, by the attractiveness of deriving even the liquid in the proto-lunar disk from the well-mixed magma ocean of Earth (Warren 1992), here we develop the chemical consequences of a different variant of the turbulent mixing hypothesis, namely, that of direct material exchange between an unstratifed magma ocean and an unstratified proto-lunar disk. In such a case, a single convective column characterizes the silicate Earth from deep regions where only liquid is present, through the critical pressure where two phases become distinguishable to the upper regions of the silicate vapor atmosphere where exchange with the melt-vapor disk presumably takes place. Similarly, a single convective column is assumed to characterize the vertical structure in the proto-lunar disk from the mid-plane up to the rarified photosphere where heat and entropy are radiated to space. Such a scenario can homogenize the relative abundances and isotopic composition of refractory elements through advection of the liquid into the atmosphere and coupling of the droplets to the vapor undergoing turbulent exchange. Furthermore, this scenario is testable because it predicts a high-level of mass-independent isotopic homogeneity in the silicate Earth-Moon system for all elements irrespective of volatility.

For the purposes of this paper, we calculate the composition at the top of the terrestrial atmosphere and assume that it is inherited by the proto-lunar disk and expressed in the lunar bulk composition without further fractionation. In the absence of phase separation, the composition of a parcel (liquid and vapor) at the top of the atmosphere reflects that of the underlying terrestrial magma ocean. Hence, such an unstratified dynamical state can generate lunar parcels isochemical with the terrestrial mantle. In the next section, we consider the consequences of partial rainout on evolving compositions distinct from the terrestrial mantle. *We note that in this paper we are only exploring chemical and isotopic fractionation in a single convective column encompassing both the magma ocean and the overlying atmosphere, and that fractionation in multiple stratified columns may have different consequences for the lunar composition.* This is a topic for future research.

In analogy with modern planetary atmospheres, we assume that the convective part of such a system is isentropic. We can make an estimate of the post giant-impact mantle entropy knowing something about the state of the pre-impact Earth and the entropy change due to the giant impact:

$$S_{final} = S_{initial} + \Delta S \tag{12}$$

where ΔS is the entropy production due to giant impact heating. We cannot know the precise state of the pre-impact Earth. However, a nearly formed Earth in the late stages of accretion will have experienced multiple giant impacts (Wetherill 1985), and cooling via solid-state convection is inefficient over the timescales of planet formation. Hence, we expect the pre-impact terrestrial mantle to be hot but at least partly solid. As a first approximation, we take the modern mantle value for the entropy of the pre-impact Earth, ~6 $k_B$/atom (Asimow et al. 2001), where $k_B$ is Boltzmann's constant. For the entropy production due to giant impact heating, we have:

$$\Delta S = \int_{T_{init}}^{T_{final}} C_P \frac{dT}{T} = \frac{3k_B}{m} \times \ln(T_{final} / T_{init}) \tag{13}$$

where m refers to the mean atomic mass (~20 amu). For initial temperatures of 2-3000 K and an average impact heating ΔT = 4,000 K, the value of the natural logarithm is nearly unity. Hence, the entropy production due to a Mars-mass giant impact is about equal to $C_p$, which is $3k_B$/atom in the high-temperature, Dulong-Petit limit. Note that we can neglect the entropy change due to melting because it is small on this scale (~0.5 $k_B$/atom). A two Mars-mass impactor would cause an average temperature increase $\Delta T$ = 8,000 K and an entropy increase of ~$5k_B$/atom. Hence, a reasonable estimate of the entropy of the post giant-impact Earth is 10 $k_B$/atom. The parcel entropy is a key parameter because it determines the thermal structure of the atmosphere, specifically, temperature (T) and vapor fraction ($f_v$) as a function of pressure. This particular value of entropy corresponds to a temperature of 3310 K and a vapor fraction of 4 percent at the reference pressure of 1 bar (although such a parcel is mostly liquid by mass, it is mostly vapor by volume and therefore dynamically behaves as a gas). However, this estimate of the entropy is an upper limit because the isentropic assumption may be inapplicable to the immediate post-impact environment, since the giant impact may preferentially deposit energy (and entropy) in the upper layers of Earth, causing an inversion layer (Canup 2004b). Some time may have to pass after the giant impact for the upper layers to cool such that calculations of a fully convective, isentropic Earth have relevance. Once the Earth becomes fully convective, however, one parameter – the specific entropy – determines the thermal structure of the silicate Earth during the course of its cooling history. With an approximate knowledge of the thermal structure, we proceed to calculate the range of compositions that can be derived from the magma ocean via rainout of droplets in the overlying silicate vapor atmosphere (Figure 1). Such a scenario is analogous to the process of condensation and rainout that results in a "dry" stratosphere (<3 ppm $H_2O$) in the modern terrestrial atmosphere.

## 6. Chemical Fractionation

Fractionation calculations in the silicate vapor atmosphere of Earth require two steps: a closed-system step, described above (section 4), whereby the liquid and vapor equilibrate at a new pressure, and an open-system step, whereby the newly-equilibrated liquid partially rains out, shifting the properties of the parcel (entropy, composition) toward that of the vapor. At each altitude, we assume that a fraction, $f_L$, of the liquid is removed via rainout. We treat $f_L$ as a free parameter, and explore the consequences of varying it for the resulting lunar composition inherited from the silicate Earth. When the liquid is partially removed, the vapor fraction in the remaining parcel increases, and is calculated according to mass balance:

$$f_{V-new} = \frac{1}{1 + (1/f_{V-old} - 1) \times (1 - f_L)} \tag{14}$$

Using the new value of the vapor fraction and equations (9) and (10), we calculate the entropy and composition of the parcel after rainout. Subsequent episodes of droplet-vapor equilibration at a lower pressure – representing adiabatic ascent – and partial liquid removal make it possible to calculate the compositional structure of the silicate vapor atmosphere undergoing varying degrees of rainout. This procedure is exactly analogous to the calculation of pseudoadiabats in atmospheric science (Holton 1992).

Plotted in Figure 2 is the result of a sample calculation with 40 percent liquid removal via rainout ($f_L$ = 0.4) every three-fold decrease in pressure (roughly every scale height), starting at a pressure of 1 bar. This particular parameter was chosen because this level of rainout over two orders-of-magnitude of atmospheric pressure can generate – at the top of the atmosphere – the widely postulated ~2x enhancement of the lunar FeO/MgO ratio.

In addition to the major elements, we are interested in following the behavior of chemical tracers during rainout. However, for this purpose, we must know the partitioning behavior of elements between liquid and vapor, which requires a solution model that yields reliable activity coefficients for trace elements at the temperatures of relevance (T = 2,500 K - 3,500 K). At present, no such solution model exists. However, we can still calculate the behavior of elements in certain limiting cases. For elements that quantitatively partition into the liquid, we have:

$$X^R_{new} = X^R_{old} \times \frac{(1-f_L)}{f_{V-old} + (1-f_L)\times(1-f_{V-old})} \tag{15}$$

where $X^R$ represents the mole fraction of the refractory tracer in the liquid before and after rainout. We use this approach to follow the fate of major-element refractory oxides (i.e. CaO, $Al_2O_3$) in the context of the rainout scenario just described. The range of lunar bulk compositions that can be evolved in this way from the Earth mantle is plotted in Figure 3. We note that compositions derived from the terrestrial mantle *enhanced* in FeO/MgO through liquid rainout are *depleted* in refractory elements. Bulk composition models featuring both an elevated FeO/MgO ratio and an enhancement in refractory element abundances in the Moon can therefore be ruled out as products of inheritance from the terrestrial mantle. In this context, it is worth noting that lunar refractory element abundances similar to (Warren 2005) or slightly depleted relative to the terrestrial mantle (Khan et al. 2007, Ringwood 1977) are commonly inferred. However, given that even the abundances of $Al_2O_3$ and CaO in the Earth's mantle are only known to within tens of percent, a precise comparison based on chemistry alone is not possible. In the next section, we discuss a method by which the lunar chemical composition can be determined with greater precision than previous techniques.

There are several parameters in these calculations, only some of which are constrained. The maximum pressure at which phase separation can, in principle, occur is dictated by thermodynamics as the pressure above which the material exists as a single phase, and is therefore well-constrained. The minimum pressure at which phase separation may occur, however, is not well known. In particular, we do not know at what atmospheric altitude exchange with the proto-lunar disk takes place (or indeed, *if* efficient exchange takes place), because it depends on the poorly understood dynamics. This is an area where numerical fluid-dynamical simulations can shed significant light on this problem. For our purposes, we consider pressure levels as low as ~10 millibars, because the silicate vapor atmosphere may be convective to such levels (Thompson and Stevenson 1988). We have taken the fraction of the liquid ($f_L$) that rains out at each altitude as a free parameter to explore the range of compositions that can be derived by fractionation from the Earth mantle, and merely note that its value may be constrained with a microphysical cloud model because only the largest droplets may have a chance to undergo rainout. We now turn to a discussion of an observational method – precise isotopic measurements – that can constrain the degree of liquid-vapor separation and, in the general framework of the scenario here explored, permit the determination of the lunar bulk composition.

## 7. Isotopic Fractionation

At sufficiently high temperature, the equilibrium isotopic differences between co-existing phases approach zero. However, small but significant isotopic fractionation is still possible at high (~$10^3$ K) temperatures if there is a change in bonding partners and/or coordination between different phases for the element under consideration. Because the vaporization of silicates involves decomposition reactions, metal-oxygen ratios of vapor species can differ from melt species equivalents, making even high temperature isotopic fractionation possible. In section 6, we explored the possibility that liquid-vapor separation in the silicate vapor atmosphere was the process responsible for any major-element differences between the silicate Earth and Moon. It is of interest, therefore, to ask what the expected range in the corresponding isotopic fractionation is and whether the bulk isotope composition of planetary bodies can be determined with sufficient precision to permit isotopic discrimination of competing scenarios.

Here, we focus on the fractionation of Si isotopes for several reasons. First, silicon in the melt presumably exists in the +4 valence state ($SiO_4^{-4}$), whereas our calculations indicate that the +2 valence (SiO) dominates the speciation of silicon in the vapor for the full range of conditions here considered, making significant (~per mil) high-temperature fractionation possible. Second, unlike the case for iron (Teng et al. 2008), oxygen (Spicuzza et al. 2007) and magnesium (Young et al. 2009), silicon isotopes have not been observed to display significant fractionation during silicate igneous processes (Fitoussi et al. 2009, Savage et al. 2010, Savage et al. 2009) making the isotopic composition of samples potentially diagnostic of the composition of source reservoirs. Third, high-precision measurements of silicon isotopes in lunar and terrestrial rocks and meteorites have been made (Fitoussi et al. 2009, Georg et al. 2007) and are ongoing (Armytage et al. 2010, Fitoussi et al. 2010). The original impetus for these measurements was the fractionation of Si isotopes due to high-pressure terrestrial core-formation. However, because Earth accretion and core formation were nearly complete by the time of the Moon-forming giant impact, and because negligible silicon partitions into the lunar core (Gessmann et al. 2001), it has become evident that the silicate Earth-Moon similarity provides evidence for the turbulent mixing hypothesis (Fitoussi et al. 2010, Georg et al. 2007). In the scenario described here, in addition to any isotopic signature inherited from the silicate Earth due to turbulent mixing, we are interested in a small mass-dependent isotopic offset that may *evolve* between the silicate Earth and Moon depending on the extent to which the liquid separated from the vapor while silicon was partially vaporized in this earliest period of Earth-Moon history.

We can estimate the magnitude of the isotopic fractionation between the vapor and liquid by considering the corresponding fractionation between vapor and crystal. The rationale behind this approach is that the bonding environment for silicon in crystalline forsterite approximates that in the liquid, consistent with the observation that suites of igneous rocks display uniform silicon isotope compositions (Fitoussi et al. 2009, Savage et al. 2010, Savage et al. 2009). To quantify the fractionation, we write the isotope exchange reaction and corresponding equilibrium constant:

$$^{30}SiO(v) + Mg_2\,^{28}SiO_4(c) <=> \,^{28}SiO(v) + Mg_2\,^{30}SiO_4(c) \qquad (16)$$

$$K_{eq} = \frac{Q(Mg_2\,^{30}SiO_4)}{Q(Mg_2\,^{28}SiO_4)} \Bigg/ \frac{Q(^{30}SiO)}{Q(^{28}SiO)} \qquad (17)$$

where (Q'/Q) is the reduced partition function ratio of the isotopically substituted and normal isotopologues in the phase of interest. For the condensate, the B-factor (or reduced partition function ratio) of silicon in crystalline forsterite has previously been calculated (Méheut et al. 2009) using density functional theory (DFT). Here we adopt the results of the DFT calculations. For silicon in the vapor, our calculations show that the SiO molecule dominates the speciation of silicon (>94 percent of silicon-bearing molecules) for the full range of conditions here encountered. We therefore approximate the silicon in the vapor with the SiO molecule and, in the standard manner (Urey 1947), calculate the ratio of reduced partition functions between isotopically substituted and normal populations assuming harmonic vibrations:

$$\frac{Q'}{Q} = \prod_i \frac{v_i'}{v_i} \frac{e^{-U_i'/2}}{e^{-U_i/2}} \frac{1-e^{-U_i}}{1-e^{-U_i'}} \qquad (18)$$

where ν (=1241 $cm^{-1}$) and ν' (=1199 $cm^{-1}$) represent the vibrational frequency for the normal and isotopically substituted isotopologue determined via spectroscopy and adopted from standard sources (Chase et al. 1985), and U is the dimensionless energy (=hν/kT). With this approximation, fractionation of $^{30}Si/^{28}Si$ between crystalline forsterite and vapor is calculated as a function of temperature. In the high-temperature limit, this fractionation can be expressed as:

$$\Delta = 5 \times \left(10^3/T\right)^2 \; per\ mil \tag{19}$$

where Δ represents the part per thousand difference in $^{30}Si/^{28}Si$ between the condensate and vapor. The isotopic fractionation here considered is strictly mass-dependent: any offset in the $^{30}Si/^{28}Si$ ratio is twice that of the $^{29}Si/^{28}Si$ ratio. Hence, we only report expected offsets for the former ratio. Equation (19) is accurate in the temperature range here encountered (T = 2,500-3,500 K) to better than ten percent. Evidently, even at the extreme temperatures encountered on the post-impact Earth, the fractionation of silicon isotopes between the liquid and vapor is large (~0.5 per mil) and can potentially generate measurable signatures (Figure 4).

To translate these calculations into a prediction for bulk silicate Earth-Moon differences, we must combine the isotopic fractionation between phases with a scenario for liquid-vapor separation. We have inferred the temperatures (section 5) and degree of phase separation (section 6) needed to derive one postulated feature of the lunar chemical composition via rainout in the silicate vapor atmosphere of Earth. We now ask what the corresponding isotopic fractionation in such a scenario would be. Since the isotopes are passive tracers, we simply adopt the rainout scenarios from the calculations above, supplementing them with statements of isotopic mass balance and fractionation:

$$\delta = f_V \delta_V + (1 - f_V)\delta_L \quad (20)$$

$$\Delta = \delta_L - \delta_V \quad (21)$$

where $f_v$ is the mole fraction of Si in the vapor, and comes from the equilibrium calculations described above (section 4), and Δ is the isotopic fractionation between the liquid and vapor and is calculated according to procedure just described (equation 19) and the calculated temperature of the parcel (section 4). As with chemical fractionation, the isotopic distillation requires two steps: a closed-system step whereby liquid and vapor isotopically equilibrate at a new temperature and pressure, and an open-system step, in which newly equilibrated liquid is partially removed via rainout, shifting the isotopic composition of the ascending parcel towards that of the vapor. An example distillation for an illustrative degree of liquid rainout ($f_L$=0.4) is shown in Figure 5. The removal of heavy-Si in liquid droplets over two orders of magnitude of pressure shifts the isotopic composition of the upper atmosphere relative to terrestrial Si. Hence, if the proto-lunar disk acquires a ~2x enhancement in FeO/MgO relative to the Earth's magma ocean via exchange with a fractionated atmosphere, a measurable isotopic offset ($\Delta^{30}$Si = +0.14 per mil) between terrestrial and lunar silicates should evolve. This is not observed (Armytage et al. 2010, Fitoussi et al. 2010, Georg et al. 2007).

Separation of silicate phases generates concomitant chemical and isotopic fractionation. However, since the degree of liquid rainout remains a free parameter, a range of outcome for the chemical and isotopic composition of the resulting Moon is possible (Figure 6). While inferring the bulk chemical composition of the Moon remains difficult and fraught with uncertainty, it has become possible to determine the isotopic composition of lunar matter to a precision that permits predictive scenarios of the lunar chemical composition to be constrained. In the absence of phase separation, the magma ocean composition is advected to high levels in the Earth's vapor atmosphere and the silicate Moon that results from exchange with such an atmosphere is isochemical with the Earth's mantle. In such a case, no isotopic offset is expected. However, *to the extent that the lunar chemical composition is distinct from the silicate Earth*, phase separation is implied and an isotopic offset between these silicate reservoirs is expected. Since no such offset has been reported (Armytage et al. 2010, Fitoussi et al. 2010, Georg et al. 2007), we cannot rule out a lunar composition that – at least in terms of major elements – is isochemical with the terrestrial mantle. Here, we merely note that more precise measurements of the lunar isotopic composition may permit a more precise determination of the lunar chemical composition.

## 8. Discussion

There is a certain similarity between the chemical composition of the Earth's mantle and the Moon. In this paper, we have explored the possibility that this circumstance has genetic significance and is due to the chemical composition of the lunar material being inherited from the partially vaporized silicate Earth in the afterglow of the giant impact. We have quantified the chemical and isotopic fractionation due to varying degrees of phase separation, and adopted the point of view that the top of the Earth's convective atmosphere is in turbulent exchange with the proto-lunar disk (Pahlevan and Stevenson 2007) as is required to explain the "terrestrial" isotopic character of lunar samples (Fitoussi et al. 2010, Georg et al. 2007, Touboul et al. 2007, Wiechert et al. 2001). To permit a comparison with observations, we have further assumed that this upper atmospheric composition is imprinted on the proto-lunar disk and expressed in the lunar bulk composition without further fractionation. This is clearly an over-simplification, as at least one subsequent stage of fractionation is required to explain the lunar volatile depletion (Ringwood and Kesson 1977), itself a fractionation event with possible isotopic consequences. Below, we discuss several avenues of research that may clarify the role of fractionation in determining the lunar composition during this earliest period of lunar history.

### 8.1 High-temperature thermodynamic model

Congruent vaporization of olivine end-members has been demonstrated experimentally at the temperatures relevant to the solar nebula ($T < 2,000$ K). However, even if such vaporization behavior persists at the higher temperatures encountered in the silicate vapor atmosphere ($T = 2,500 - 3,500$ K), a two-component liquid model must be considered a crude approximation to the multi-component solution of the terrestrial magma ocean. More realistic solution models would permit several important questions to be addressed. First, aside from the FeO/MgO ratio, other chemical variables may undergo changes due to phase separation; in particular, Si/(Mg+Fe) fractionation may take place. Such fractionation is rendered more likely in the silicate vapor atmosphere because, in addition to olivine, the Earth's mantle – the ultimate source of the vapor atmosphere – contains plenty of pyroxene and enstatite has been experimentally shown to vaporize *incongruently*, with the vapor enriched in silica (Mysen and Kushiro 1988). Therefore, rainout on the post-impact Earth may have resulted in a certain degree of Si/(Mg+Fe) fractionation that would be expressed in an olivine/pyroxene ratio for the lunar mantle that is distinct from the silicate Earth, a feature that is, in fact, found in certain lunar composition models (Ringwood 1977). Second, robust articulation of the isotopic consequences of phase separation depends on such factors as the atmospheric thermal structure, the elemental partitioning, and the vapor speciation, all of which depend on an accurate thermodynamic description at the relevant temperatures. In addition to refining the predictions for silicon isotopes, a high-temperature solution model would permit robust calculation of the consequences of phase separation for the isotopic composition of oxygen, magnesium, and iron, elements whose isotopic abundances in planetary materials continue to be determined with increasing precision (Bourdon et al. 2010, Liu et al. 2010).

### 8.2 Volatile Trace Elements (Na, K, etc)

The single-stage scenario explored in this paper is one where the proto-lunar disk acquires its chemical composition from the terrestrial magma ocean and, depending on the degree of rainout in the intervening atmosphere, is variously *enhanced* in its abundance of volatile elements relative to the silicate Earth. This is contrary to the observations. In particular, elements that condense from the solar nebula at temperatures below about 1,200 K are – with only a few exceptions – *depleted* in lunar samples, as observed in the lunar basalts (Ringwood and Kesson 1977). Therefore, the scenario explored here is, at best, an incomplete story. Such a scenario can be reconciled with the lunar volatile depletion by considering additional episodes of fractionation during lunar formation distinct from that which occurred in the silicate vapor atmosphere of Earth. In particular, the depletion of volatile trace elements may be due to disk processes, such as gravitational separation of droplets to the mid-plane (Machida and Abe 2004) and the radial transport of volatile elements to the Earth, a process analogous to the turbulent redistribution of water vapor beyond the ice line in the solar nebula (Stevenson and Lunine 1988). Other possibilities include an episode of volatile loss from the proto-lunar disk via escape in a hydrodynamic wind (Genda 2004, Genda and Abe 2003) or the separation and loss of volatiles after the disk phase, during the process of lunar accretion (Abe et al. 1998), or even outgassing from a naked magma ocean followed by escape of the transient vapor atmosphere. For the scenario presented here to have predictive power in terms of lunar composition, the episode of volatile element depletion must have been divorced from any major-element fractionation. Whether or not this was indeed the case may be determined through physical and chemical modeling of liquid-vapor fractionation in each of these various stages of lunar formation.

### 8.3 The Role of Metals

We have thus far assumed that the Earth's metallic core – a liquid denser than, and immiscible with, the overlying magma ocean – was isolated during and after the turbulent mixing process and that metals did not play a role in lunar formation. Such a scenario does not explain the probable existence of a small lunar core (Konopliv et al. 1998). The conventional scenario for the origin of this putative core is that the predominantly silicate disk that was injected into orbit by the giant impact contained a few percent metal from the core of the impactor. However, in the context of the unstratified mixing scenario here explored, both the core and the metal/silicate ratio of the Moon were inherited from the terrestrial magma ocean while metallic droplets were held in suspension. This idea is rendered more likely by the fact that a significant fraction of the cores of Earth-forming impactors emulsified into droplets and equilibrated with the silicate Earth upon impact (Nimmo et al. 2010), and that metallic droplets, once generated, remain suspended in the convective magma ocean for a timescale comparable to the turbulent mixing timescale (Stevenson 1990). In terms of observable signatures, any such "last gasp" of terrestrial core segregation that postdates turbulent exchange with the disk might alter the story here described in two ways. First, if significant amounts of FeO partition into the core-bound metal (Rubie et al. 2004), then the FeO/MgO value for the modern mantle would not reflect that of the magma ocean in the afterglow of the giant impact. Quantitative scenarios seeking to derive the lunar composition from the silicate Earth would have to take such effects into account. Second, if a significant amount of silicon dissolves into these last metallic droplets (Gessmann et al. 2001), core segregation can evolve an isotopic offset between the silicate Earth and Moon even if the system previously underwent isotopic homogenization. At present, the sign, but not the magnitude of such isotopic fractionation is known, and it is the same sign as the offset expected due to liquid-vapor fractionation in the vapor atmosphere, with the heavy isotopes, in both cases, being concentrated in the silicate Earth. This circumstance is fortuitous, because it means that the absence of an isotopic offset cannot be explained by cancellation of the effects of liquid-vapor separation followed by terrestrial core segregation, and precise measurement of silicon isotopes limits the efficacy of each of these processes in turn.

## 9. Conclusions

The turbulent mixing hypothesis presents us with an opportunity. Progress in developing and testing the giant impact hypothesis has been hampered by an inadequate knowledge of the composition of the impactor that triggered lunar formation. That the lunar isotopic composition is identical with the terrestrial mantle to high precision for a diverse range of tracers argues for the possibility that the lunar material was processed through the Earth's post-impact vapor atmosphere and underlying magma ocean. In such a scenario, and without liquid-vapor separation, the silicate portion of the Moon would be expected to be isochemical with the Earth mantle. Therefore, observed chemical differences in the silicate Earth-Moon system demand explanation. Here, we have explored a scenario in which chemical fractionation in the silicate vapor atmosphere of Earth is solely responsible for any major-element differences in the silicate Earth-Moon system. Our results demonstrate that an enhancement – relative to Earth mantle – of the FeO/MgO ratio must be accompanied by a depletion of refractory element abundances, ruling out certain lunar bulk composition models. Furthermore, chemical and isotopic fractionations are concomitant. The role of silicate vapor petrology in determining the lunar chemical composition can therefore be constrained via precise measurement of isotopes that fractionate at high temperatures. Due to its strikingly different bonding environment in the liquid and vapor, the isotopes of silicon provide a sensitive tracer of fractionation in the afterglow of the giant impact. Existing measurements of the isotopes of silicon in terrestrial and lunar silicates thus far yield no evidence for a lunar FeO/MgO ratio that is distinct from the terrestrial mantle. However, accurate predictions of isotopic fractionation require accurate knowledge of the partial vaporization behavior of the element under consideration. Hence, in addition to precise isotopic measurements, progress in developing the lunar origin scenario requires better thermodynamic models of silicate liquids at the temperatures encountered during lunar formation. The scenario explored here is a first step towards understanding the role of liquid-vapor fractionation in imprinting the cosmochemical signatures of the giant impact.

## Acknowledgements

We would like to thank Edwin Schauble for sharing detailed results of his calculations, Nicolas Dauphas and Paul Asimow for helpful discussions and comments on the manuscript, and Paul Warren and an anonymous reviewer for thorough reviews that greatly improved the manuscript.

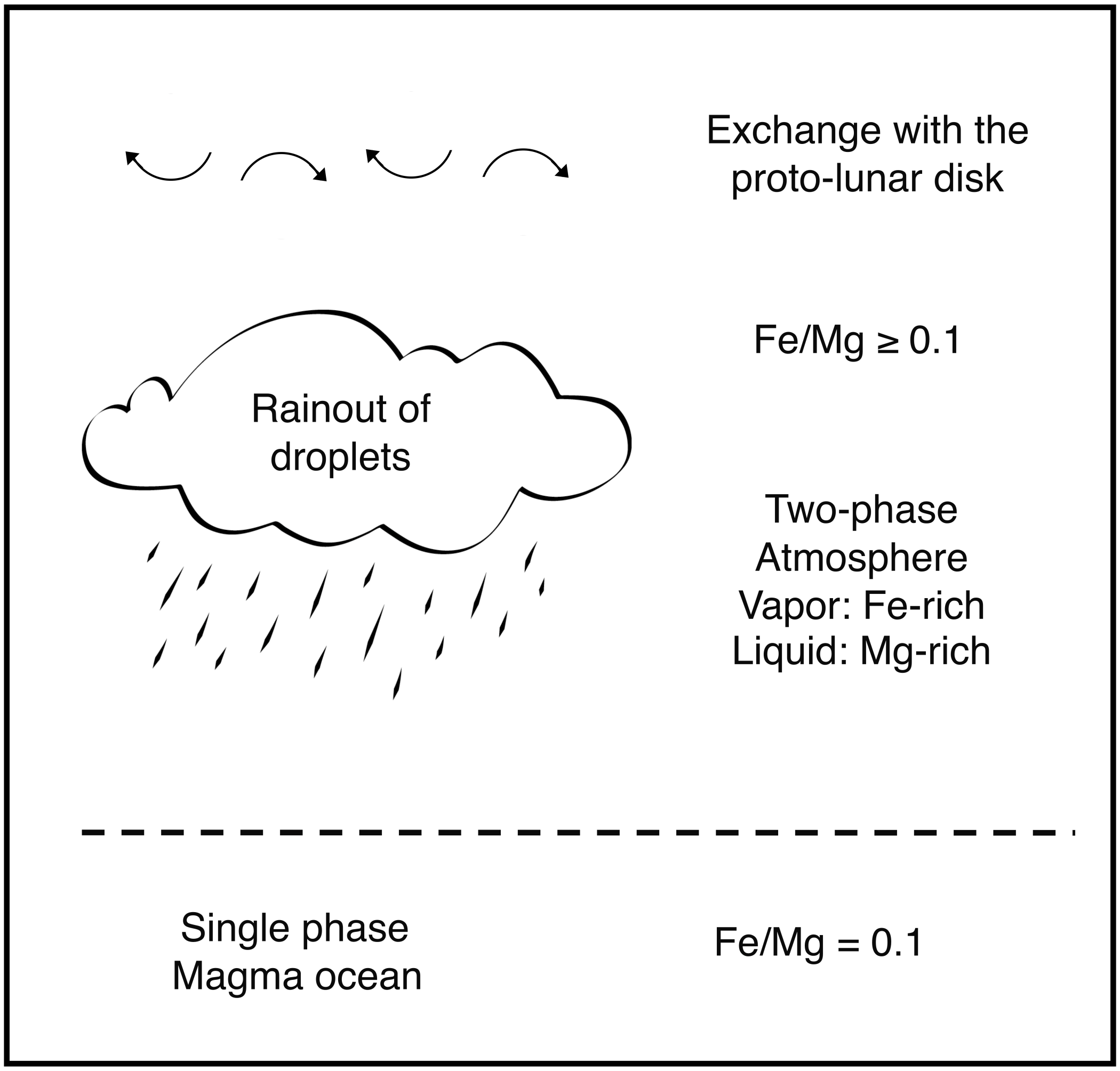

**Figure 1.** Chemical fractionation on an unstratified Earth. A single convective column characterizes the Earth from the deep magma ocean, where only one phase is present, through the top of the two-phase atmosphere. Rainout of Mg-rich droplets in ascending parcels shifts the composition of the upper atmosphere towards an Fe-rich vapor composition. The degree of rainout in these models is a free parameter.

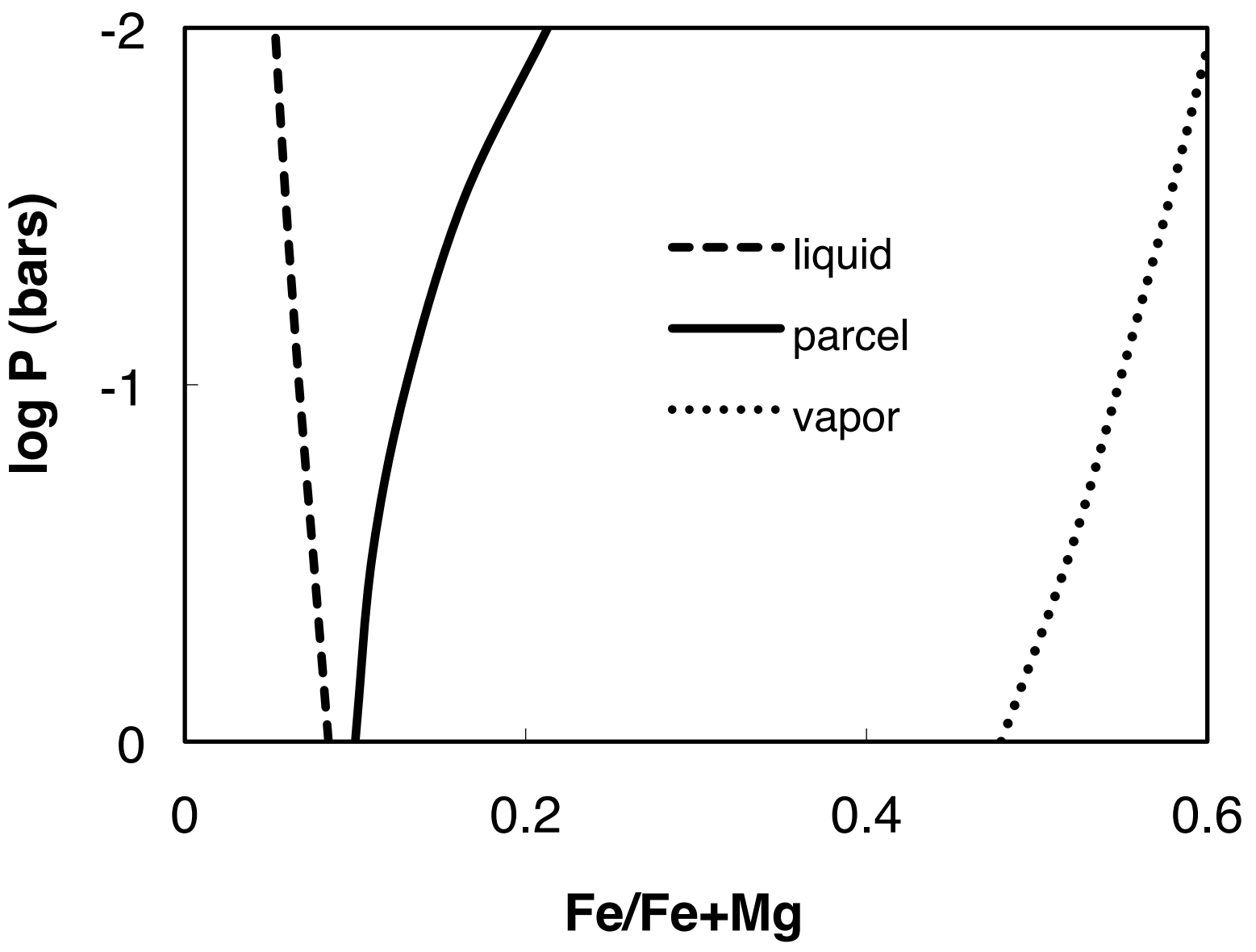

**Figure 2.** Chemical structure of the silicate vapor atmosphere of Earth undergoing rainout. The parcel represents the composition of the atmosphere (forsteritic droplets suspended in a fayalitic vapor) and shifts with altitude toward that of the Fe-rich vapor as the droplets separate via rainout. The lower atmosphere is directly sourced by convection from the underlying magma ocean and is therefore of the same composition. This calculation assumes that 40 percent of the liquid is removed via rainout every three-fold decrease in pressure ($f_L$ =0.4). This parameter is chosen such that – at the top of the atmosphere – a two-fold enhancement in the FeO/MgO ratio evolves. This enhancement is comparable to a widely postulated silicate Earth-Moon difference, and has observable consequences (see text).

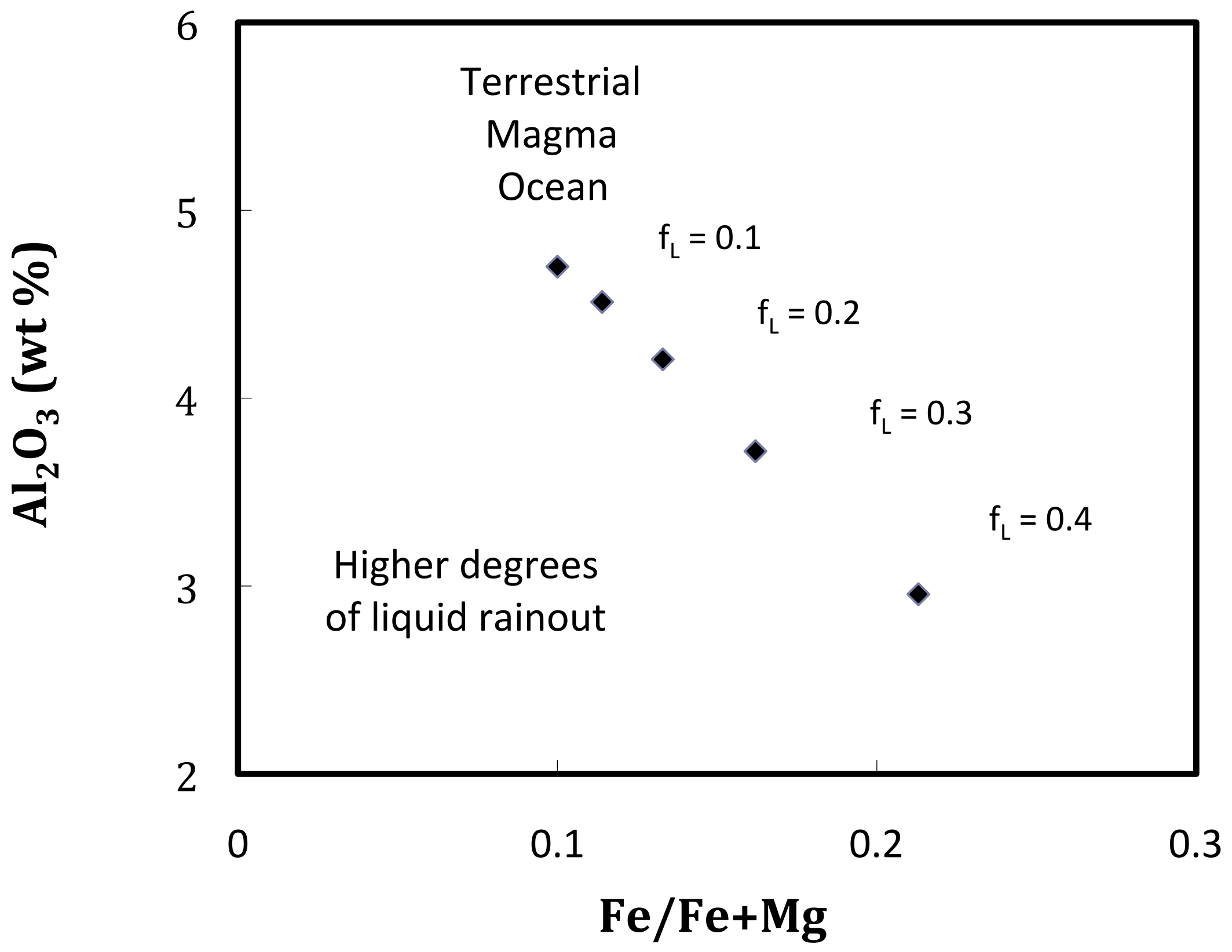

**Figure 3.** Co-evolution of the Fe/Fe+Mg ratio and refractory element abundances – represented by alumina – of atmospheric columns undergoing rainout. The alumina abundance (=4.7 weight percent) and Fe/Fe+Mg (=0.1) of the modern mantle is assumed for the magma ocean. Plotted are upper atmospheric compositions evolved by rainout at lower levels and inherited by the proto-lunar disk. The process of phase separation in the vapor atmosphere shifts the expected lunar composition towards the composition of the vapor with an enrichment in Fe/Fe+Mg and depletion in refractory elements, but small degrees of phase separation can yield lunar compositions nearly isochemical with the terrestrial mantle.

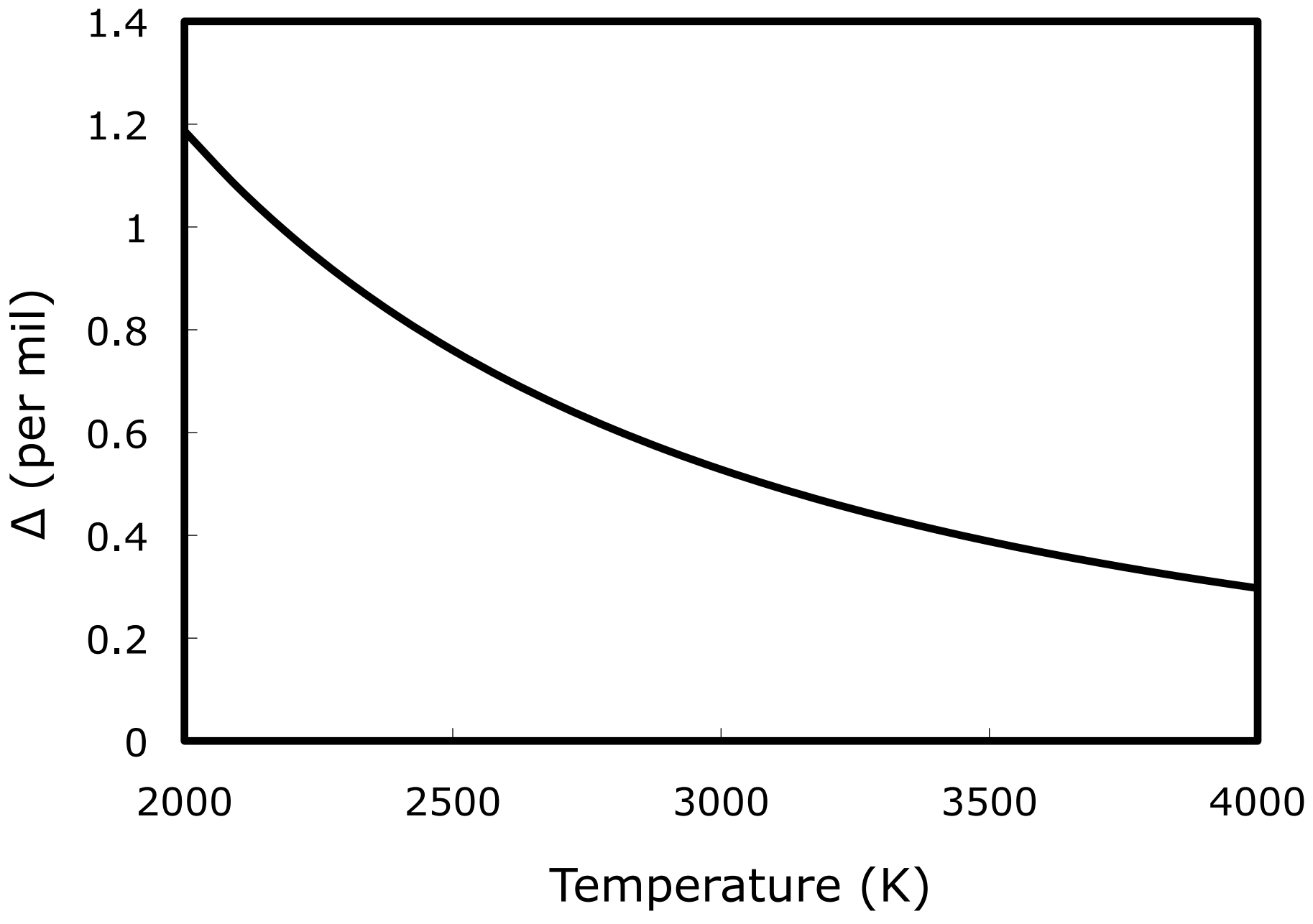

**Figure 4** – The fractionation of $^{30}Si/^{28}Si$ between the liquid and vapor, assuming that the isotopic composition of crystalline forsterite represents the isotopic composition of silicon in the liquid. $\Delta = (K_{eq}-1) \times 10^3$ for values of K near unity, expressing the difference in $\delta^{30}Si$ between liquid and vapor. According to this calculation, measurable isotopic fractionation is possible, even at the high temperatures encountered on the post-giant impact Earth (2-4000 K).

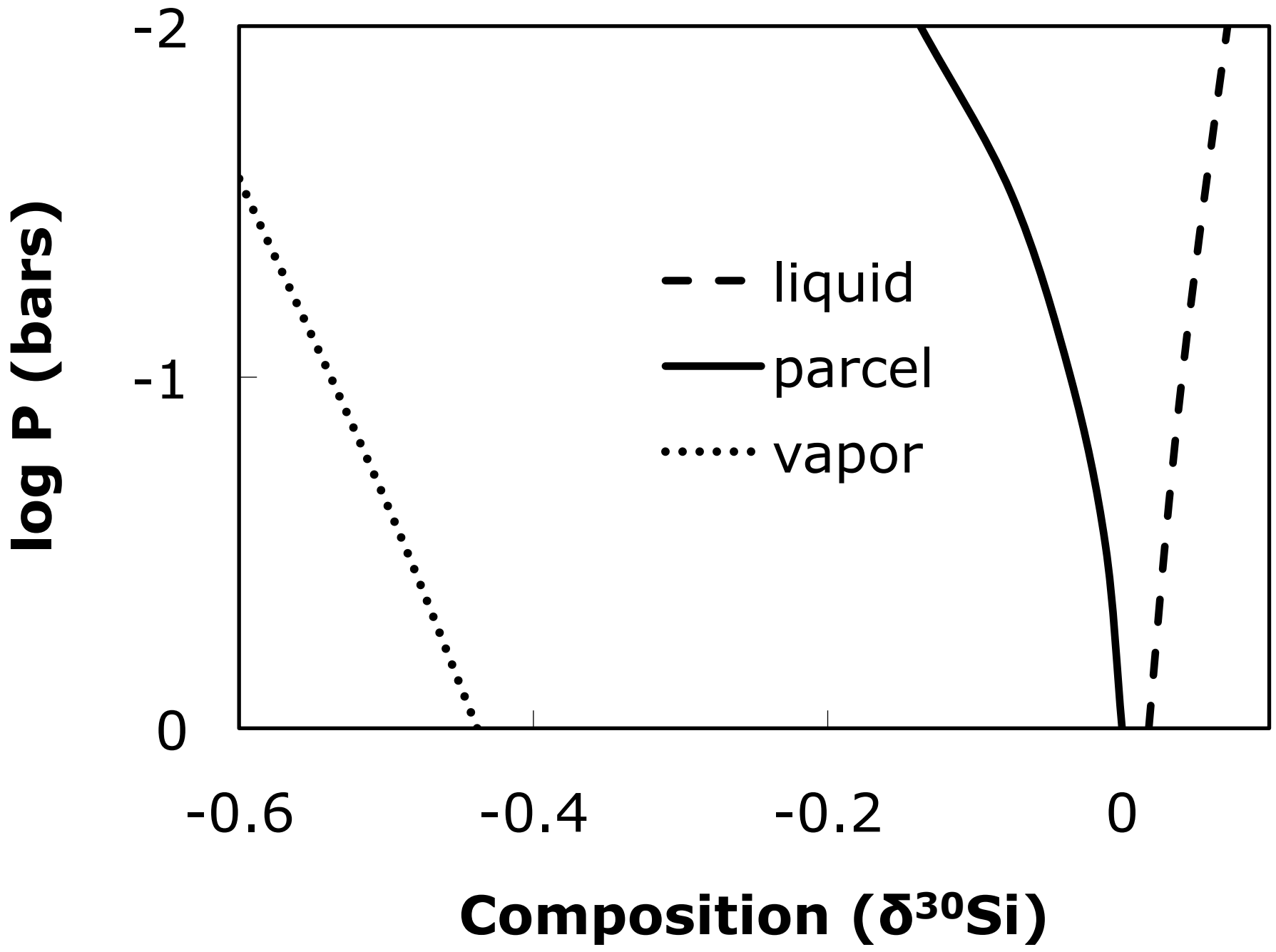

**Figure 5.** Isotopic structure of the silicate vapor atmosphere of Earth undergoing the same degree of rainout as Figure 2 ($f_L = 0.4$). Removal of isotopically heavy liquid during ascent shifts the composition of the upper atmosphere towards lighter compositions. The liquid-vapor separation occurs at temperatures of 2,500-3,500 K. If rainout throughout the column was the process responsible for a two-fold enhancement in the lunar FeO/MgO ratio, a ~0.14 per mil offset in $^{30}Si/^{28}Si$ in the silicate Earth-Moon system should have evolved. This is not observed.

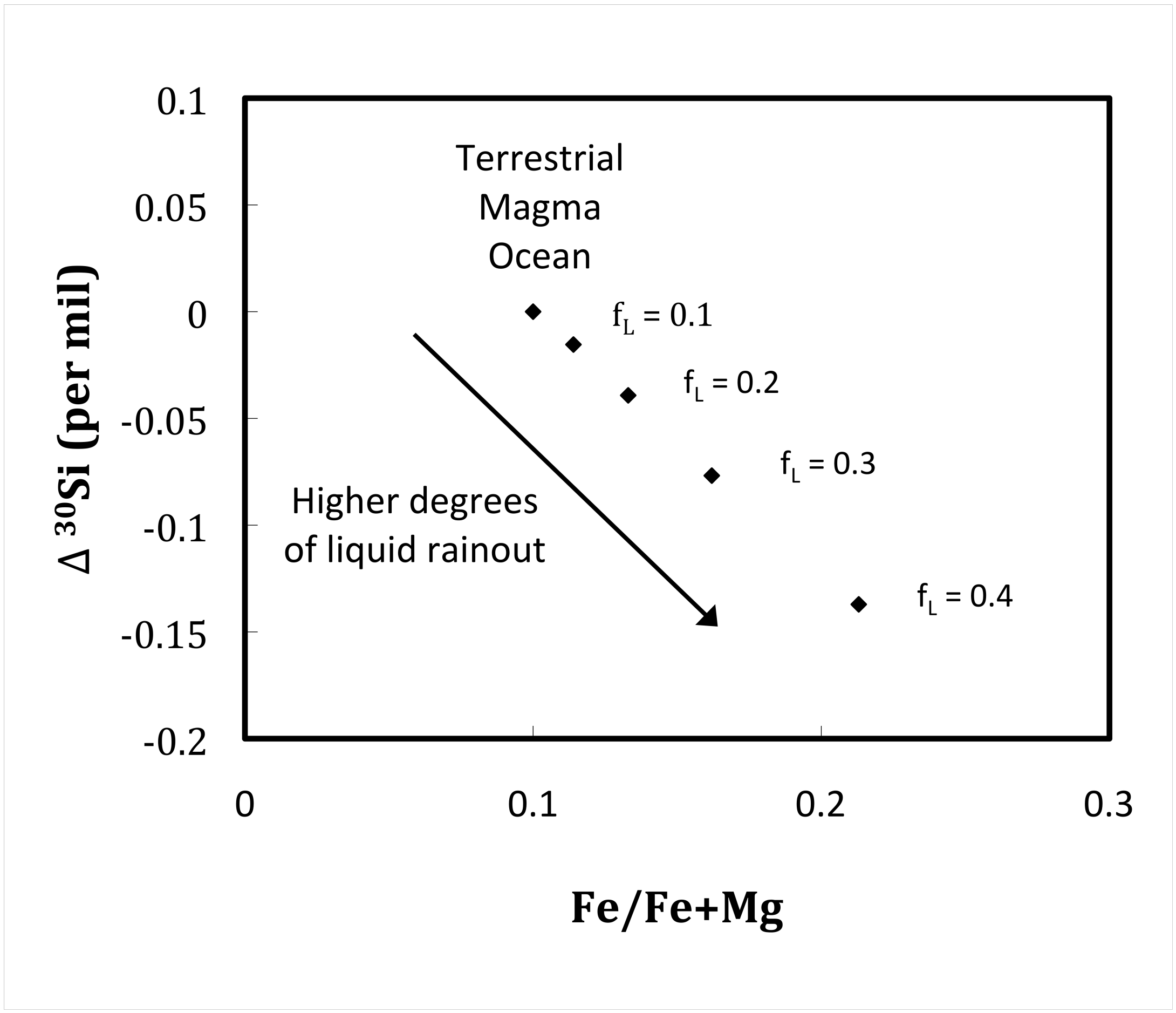

**Figure 6.** Co-evolution of chemical and isotopic compositions in atmospheric columns undergoing various degrees of rainout. Plotted are upper atmosphere compositions inherited by the proto-lunar disk. $\Delta^{30}$Si represents the part per thousand difference in the $^{30}$Si/$^{28}$Si ratio between the magma ocean and upper atmosphere (=$\delta^{30}Si_M$-$\delta^{30}Si_E$), the expected difference between the silicate Earth and Moon. Since both the FeO/MgO ratio and the isotopes of silicon are fractionated via liquid rainout, precise silicon isotope measurements can place precise constraints on the chemical composition of the bulk Moon. Isotopic measurements represent a new method by which to determine the lunar bulk composition.

**Table 1** – The standard entropy and enthalpy of reactions (ln K = $\Delta G^o/RT$, $\Delta G^o = \Delta H^o - T\Delta S^o$) that determine the vapor pressure and speciation in equilibrium with liquid olivine.

| Reaction | $\Delta S^o$ (J/mol.K) | $\Delta H^o$ (kJ/mol) | Reference |
|---|---|---|---|
| $Mg_2SiO_4$ (l) = 2 MgO (v) + $SiO_2$ (v) | 490 | 1888 | 30 |
| $Fe_2SiO_4$ (l) = 2 FeO (v) + $SiO_2$ (v) | 412 | 1500 | 30,31 |
| MgO (v) = Mg (v) + ½$O_2$ (v) | 24.1 | 77.4 | 30 |
| FeO (v) = Fe (v) + ½$O_2$ (v) | 52.6 | 174 | 30 |
| $SiO_2$ (v) = SiO (v) + ½$O_2$ (v) | 76.6 | 197 | 30 |
| SiO (v) = Si (v) + ½$O_2$ (v) | 67.7 | 557 | 30 |
| $O_2$ (v) = 2 O (v) | 134 | 509 | 30 |

**Table 2** – Entropy data for pure liquids and vapors at 1 bar (S = A.lnT + B).

| Substance | A (kJ/mol.K) | B (kJ/mol.K) | Reference |
|---|---|---|---|
| $Mg_2SiO_4$ (l) | 189 | -992 | 30 |
| $Fe_2SiO_4$ (l) | 192 | -905 | 30,31 |
| $SiO_2$ (v) | 59.8 | -120 | 30 |
| SiO (v) | 36.8 | -2.1 | 30 |
| Si (v) | 22.1 | +41.5 | 30 |
| MgO (v) | 48.5 | -51 | 30 |
| Mg (v) | 21.7 | +23.8 | 30 |
| FeO (v) | 38.5 | +17.9 | 30 |
| Fe (v) | 25.2 | +35.7 | 30 |
| $O_2$ (v) | 37.5 | -15 | 30 |
| O (v) | 21.2 | +40.1 | 30 |